\documentclass[aps,prd,twocolumn,groupedaddress,amsfonts,amssymb,amsmath,showpacs,showkeys]{revtex4-1}
\usepackage{graphicx}

\usepackage{natbib}

\newcommand{\jcap}{Journal of Cosmology and Astroparticle Physics}

\newcommand{\mnras}{MNRAS}

\begin{document}

\title{MOdified Newtonian Dynamics as an entropic force}
\author{D.A. Carranza}
\email[Email address: ]{dortiz@astro.unam.com}
\author{S. Mendoza}
\email[Email address: ]{sergio@astro.unam.mx}
\affiliation{Instituto de Astronom\'{\i}a, Universidad Nacional
                 Aut\'onoma de M\'exico, AP 70-264, Distrito Federal 04510,
	         M\'exico \\
            }


\begin{abstract}
Under natural assumptions on the thermodynamic properties of space and
time with the holographic principle we reproduce a MOND-like behaviour
of gravity on particular scales of mass and length, where Newtonian
gravity requires a modification or extension if no dark matter component
is introduced in the description of gravitational phenomena.  The result
is directly obtained with the assumption that a fundamental constant
of nature with dimensions of acceleration needs to be introduced
into gravitational interactions. This in turn allows for
modifications or extensions of the equipartion law and/or the holographic
principle.  In other words, MOND-like phenomenology can be reproduced when
appropriate generalised concepts at the thermodynamical
level of space and/or at the holographic principle are introduced.
Thermodynamical modifications are reflected in extensions
to the equipartition law which occur when the temperature of the system
drops below a critical value, equals to Unruh's temperature
evaluated at the acceleration constant scale introduced for the description
of the gravitational phenomena. Our calculations extend the ones by
\citet{verlinde} in which Newtonian gravity is shown to be an emergent
phenomenon, and together with it reinforces the idea that gravity at
all scales is emergent.
\end{abstract}


\maketitle

\section{Introduction}
\label{introduction}

\noindent The laws for black hole mechanics have 
suggested a remarkable similarity with the three laws of thermodynamics,
in such a way that quantities associated to black hole properties have
their corresponding thermodynamic equivalent interpretation 
\citep{bekenstein73, bekenstein74, hawking73}.
In particular, the black hole area --which is determined
by its horizon-- is related to the associated black hole entropy, in
the sense that it cannot decrease in time under any physical process on a
closed system. The temperature of the black hole is given by the
Hawking-Zeldovich temperature and is inversely proportional to the mass of
the black hole~\citep{hawking74,townsend}.
The well known interpretation of entropy
as a quantity that offers a measure of non-available information --or
disorder-- in a system, has leaded directly to the idea that the increase
in entropy, and therefore in area, is due to the loss of information.  This
is due to the fact that when a particle crosses the horizon it has no 
more causal relation with the rest of the universe \citep{bekenstein73}.

All the above suggest the possibility for a deep relation between
thermodynamics and gravity. This has been studied mainly in the
relativistic regime under the concept of emergent gravity, considering
thermodynamics as a more fundamental theory from which, general relativity
can be derived (see e.g.~\cite{padma_review} and references therein).
Using a metric treatment of the thermodynamic variables in a curved
space-time, \cite{jacobson95} has been able to derive Einstein's field
equations. In the non-relativistic regime, \cite{verlinde} used very
simple assumptions about space, energy and information in order to show
that the first law of thermodynamics, along with an entropy formula,
leads directly to Newton's law of gravity. 

  In recent years, a growing number of independent observations
have suggested that gravity requires modification \cite{hernandez12,
hernandez12a, mastropietro, lee, thompson, moffat} and not the inclusion
of unknown dark matter entities.  All these facts are encouraging, since
similar arguments like Verlinde's ones can be used to search for a more
profound fundamental basis for an extended theory of gravity like the
one proposed by \cite{mendoza11}. This approach is based on
Buckingham's $\Pi$-Theorem of dimensional analysis, and deals with the
problem of a test particle under a gravitational field generated by a
central point mass $M$ in an extended regime of gravity.  Such analysis
gives the general form for the acceleration $a$ experienced by a test
particle at a distance $r$ from the central mass:

\begin{equation}
 a = a_0 f(x), \quad \textrm{with} \quad
 x:=\frac{l_M}{r}:=\frac{1}{r}\left(\frac{GM}{a_0}\right)^{1/2},
\label{accel_mendoza}
\end{equation} 

\noindent where $a_0$ is Milgrom's acceleration constant
\cite{milgrom83c}, introduced as an extra fundamental constant of nature
when extended gravitational phenomena is described.
The unknown dimensionless function $f(x)$ depends only on the dimensionless 
quantity $x$, which according to the $\Pi$-Theorem, is the key parameter in
a description of extended gravity. The 
function \( f(x) \) is such that: 
$f(x) \rightarrow x^2$ when \( x \gg 1 \), converging to Newtonian gravity;
and $f(x) \rightarrow x$ when \( x \ll 1 \), reaching the deep MONDian
regime of gravity.  The authors also showed that the transition function
can be written as:

\begin{equation}
 f(x) = x \frac{1\pm x^{n+1}}{1\pm x^n},
\label{f_series}
\end{equation} 

\noindent where $n$ is a constant that must be fixed 
via astronomical observations.  A large value of \( n \) means that the function
\( f(x) \) abruptly changes from a MONDian to a Newtonian regime of gravity
about \( x \approx 1 \).  Small values of \( n \) yield soft transitions
about the same point.

  For the case of spherically symmetric mass distribution,
this extended Newtonian gravity approach proposal reproduces a MOND-like phenomenology \cite{milgrom83c}
and has proven to be in good agreement with observations in astrophysical
systems across different scales without invoking any dark matter component
\cite{bernal11,carranza13,hernandez12,hernandez12a,mendoza13,mendoza11}.

 In another attempt to obtain the MOND-like force formula, \cite{debye_article}
have followed Verlinde's work assuming that the equipartition
law of energy is modified. In this approach, a Debye's function is introduced, 
and becomes identified as the MOND interpolation function $\mu(a/a_0)$.
The reason of doing so is simply to satisfy Newtonian and MONDian regimes
of gravity at their corresponding limits. 
Furthermore, they also give an expression for the value of Milgrom's
acceleration constant in terms of Debye's temperature $T_\textrm{D}$:

\begin{equation}
a_0 = \frac{12ck_\textrm{B}T_\textrm{D}}{\pi \hbar},
\label{a0_debye}
\end{equation} 

\noindent thus interpreting it as a cut off temperature below which,
modifications to the equipartition law must occur. This approach takes
into account the fact that the dynamical sector needs to be modified
and so, the validity of Newton's law of gravitation remains unaffected.
As explained by the dimensional analysis of \cite{mendoza11}, the
modification must occur in the gravitational sector and rather than
working with a transition function \( \mu(a/a_0) \), the extension is
carried out through the inclusion of a transition function \( f(x) \).

 In this work, we show how, using arguments about thermodynamics and
information, it is possible to derive in several ways an equation for
the gravitational force in an extended modified gravity regime, which
supports the idea that gravity can be understood as an emergent force,
i.e. a consequence of deeper fundamental principles. The article is
organised as follows: in section \ref{mod_equip} we review the main
hypothesis made by \cite{verlinde}, and then we use dimensional analysis
arguments to study modifications to the equipartition law of energy via
two approaches, one of them purely thermodynamic, and the other only
gravitational. Using another point of view, section~\ref{mod_information}
is devoted to study possible modifications to the holographic principle,
specifically, to the number of bits contained inside a screen under the
assumption that Milgrom's acceleration constant $a_0$ is a fundamental
constant of nature. Finally, in section \ref{discussion} we discuss our
main results.

\section{Modifications to the equipartition law}
\label{mod_equip}

 \noindent
We begin this section reviewing briefly some of the main ideas and hypothesis
made in Verlinde's work \cite{verlinde}. One of the most important assumptions
made by him, is that the information describing a physical system is stored
on spatial surfaces, or \emph{screens}, that are ruled by the holographic
principle. Every surface behaves as a ``stretched horizon'' of a black hole
(although in this case it has no physical properties like density or surface
pressure), and when a particle interacts with it, the entropy, and consequently,
the amount of information gets affected. In principle we do not know the
shape of the surfaces, so for simplicity we can consider each screen as
closed and spherical with  radius \( r \). Each surface contains $N$ bits
of information. One can also think that on each fundamental Planck square
area, the maximum information that can be stored is one bit. This length
is constructed with three fundamental constants of nature: (1) Newton's
constant of gravity $G$, (2) the velocity of light $c$ and (3) Planck's
constant $\hbar$. With these assumptions, the number of bits $N$ stored on
each screen can be expressed as:

\begin{equation}
 N = \frac{A}{l_\textrm{P}^2} = \frac{4\pi r^2}{l_\textrm{P}^2},
\label{nbits}
\end{equation} 

\noindent being $A$ the area of the spherical screen and $l_\textrm{P}^2=
G\hbar /c^3$ the Planck area.

 The main motivation by \cite{verlinde} to think of gravity as a force
related to entropy has its origin on the restitutive force that acts on
a polymer when it suffers a displacement $\Delta x$. This force tends
to restore the polymer to its original position since this configuration
maximises the entropy. The link with gravitation consists on a similar
idea, for which there is an entropic force that emerges as a consequence
of the system searching for a configuration of maximum entropy when a
particle approaches a given screen. We assume that inside the
screen the dynamics allow us to define energy, and consequently the
associated mass $M$ and temperature $T$ are well defined quantities. With
this, we can use the first law of thermodynamics to find the force $F$
associated to changes in the stored information, i.e. due to a change in
entropy $\Delta S$ given by:

\begin{equation}
  F\Delta x = T\Delta S,
\label{thermo}
\end{equation}

\noindent for a constant volume. Let us now find the expression for the
gravitational force by considering gravity as an entropic force. For this,
we follow the approach by \cite{verlinde} and \cite{jacobson95} analysing
the behaviour of a test mass \( m \) particle near a black hole horizon. At a
distance of one Compton length from the horizon, the particle can be considered
to be part of the black hole and so, its entropy is increased in the following
way:

\begin{equation}
  \Delta S = 2\pi k_B\frac{mc}{\hbar}\Delta x,
\label{entropy}
\end{equation}

\noindent when a displacement $\Delta x$ occurs.  In other words, a change
in the particle position causes an increment on the system's entropy, such
that it tries to maximise it and as such, the horizon can be substituted
by a screen. The other key assumption to make is that the energy contained
inside the surface satisfies the principle of equipartition, and that it
is equal to $M c^2 $, where \( M \) is its associated mass. At this point,
we take an approach similar to the one followed by \cite{debye_article},
i.e. we search for a modification of the equipartition law. This can be
achieved with the help of Buckingham's theorem of dimensional analysis,
since it provides a way to find the general form for the energy when
modifying effects are considered.

 To do so, note that the dimensional relevant quantities  of the problem
are the energy $E$ associated to the screen, it's temperature \( T \)
--or more important for a dimensional analysis treatment its energetic
temperature \( k_\text{B} T \) (where \(k_\text{B} \) is Boltzmann's
constant), the speed of light $c$ and Planck's constant $\hbar$.  Since we
are studying a gravitational problem in an extended regime, Milgrom's
constant $a_0$ is also introduced, along with Newton's constant of
gravity $G$.  The associated mass \( M \) of the spherical screen
with radius \( r \) is given by \( E = M c^2 \).  In other words, seven
physical quantities (\( k_\text{B}T,\ c,\ \hbar,\ G,\ a_0,\ M,\ r\)) play
a fundamental role in the description of the energy \( E \) of a screen.
Since there are three independent dimensions (length, time and mass),
Buckingham's \( \Pi \) theorem demands the energy of the screen to be
given by \cite{sedov}:

\begin{equation}
 E = \frac{1}{2}Nk_\textrm{B}T F \left(x, \frac{T}{T_*},
     \frac{\lambda}{r}, \frac{l_\textrm{P}}{r} \right),
\label{energy_general}
\end{equation}

\noindent where $F$ is an unknown function of four dimensionless
parameters, $x$ was defined in equation ~\eqref{accel_mendoza}, $T_* :=
a_0 \hbar/ck_\textrm{B}$ is a quantity with dimensions of temperature 
-which coincides to the Unruh temperature evaluated at the acceleration
\( a_0 \)- and $ \lambda := c^2/a_0 $ is a characteristic length that
appears when relativistic effects are considered in the extended regime
of gravity \cite{bernal11}. The dimensionless factor $N/2$ has been
introduced for consistency with the standard law of equipartition.  It is
important to note that the temperature $T_*$ is proportional to the so
called ``cut off Debye temperature'' given in equation~\eqref{a0_debye}
and constitutes a characteristic temperature scale in an extended regime
of gravity where MONDian effects are to be taken in consideration.

  The ratio $l_\textrm{P}/r$ appears as an characteristic dimensionless 
quantity in equation~\eqref{energy_general}, but equation \eqref{nbits}
suggests that this quantity is closely related to the holographic principle 
and not to a modification to the equipartition law.  Also, since that ratio
does not contain \( a_0 \) it will be incapable to account for any MOND
phenomenology.  In other words, the equipartition energy \( E \) is only a
function of three dimensionless parameters: \( x,\ T/T_* \) and \( \lambda
/ r \).  The simplest assumption to make for the function \( F \) is that
it is of power-law form on any of its arguments, i.e.:

\begin{equation}
 E = \frac{\xi}{2}Nk_\textrm{B}T \, x^\alpha \, 
     \left( \frac{T}{T_*} \right)^\beta \, \left(  \frac{\lambda}{r}
     \right)^\gamma,
\label{energy_general-2}
\end{equation}

\noindent where \( \xi \) is a constant of proportionality.

  In the remaining of this section, we study three separate cases
associated with the previous relation:

\subsection*{Case (A)}
\label{subsection1}

 Let us consider \( \alpha=\gamma = 0\) in
equation~\eqref{energy_general-2} to obtain:

\begin{equation}
 E = \frac{\xi}{2}Nk_\textrm{B}T
     \left(\frac{T}{ T_* }\right)^\beta.
\label{enregy1}
\end{equation}

\noindent The physics behind this choice of parameters can be understood
under the basis of the case studied by \cite{debye_article} since, as it was
mentioned previously, the temperature $T_*$ corresponds -except  for a
$2\pi$ factor- with the Unruh temperature on a holographic screen. In that
particular study, the authors dealt with a non--standard equipartition law
of energy including the one dimensional Debye function to take into account
the corrections at low temperatures. Their equipartition law corresponds
to \eqref{enregy1}, with \( T \) representing the one dimensional Debye 
function.

\noindent  Equating relation~\eqref{enregy1} to $Mc^2$, and using 
equation~\eqref{nbits}, the temperature can be written as:

\begin{equation}
 T = \frac{\hbar}{ck_\textrm{B}}\left(\frac{a_0^\beta G M }{ 
   2\pi \xi r^2}\right)^{1/(\beta+1)}.
\label{temp1}
\end{equation}

\noindent Substituting this into \eqref{thermo}, and employing \eqref{entropy},
as made by Verlinde, the resultant entropic force is:

\begin{equation}
 F = 2\pi m \left(\frac{a_0^\beta G M }{2\pi \xi_1 r^2} \right)^{1/(\beta+1)}.
\label{force1}
\end{equation}

  Taking \( \beta = 0 \) and \( \xi = 1 \) in equation~\ref{enregy1}, the
standard equipartition law is obtained and as seen from
relation~\ref{force1}, Newton's gravitational law is recovered.
In order to obtain a MONDian \( 1 / r \) gravitational force law we must
take $\beta = 1$ and $\xi=2\pi$:

\begin{equation}
 F = m \frac{\sqrt{a_0GM}}{r}.
\label{force1a}
\end{equation}

  In other words, the equipartition energy must satisfy the following
condition:

\begin{equation}
  E = \frac{1}{2}Nk_\textrm{B} T
      \begin{cases}
        1, & \text{for Newtonian gravity,} \\
	2\pi T/T_*, & \text{for MOND-like gravity.}
      \end{cases}
\label{energycases1}
\end{equation}

\subsection*{Case (B)}
\label{subsection2}

  Let us now consider the case when \( \beta = \gamma = 0 \) in
equation~\eqref{energy_general-2}.

 Once again, using the equivalence
between mass and energy, and the explicit form of Planck's area, we can
write:

\begin{equation}
 Mc^2 = \xi \frac{2\pi r^2 c^3}{G\hbar}k_\textrm{B} T x^\alpha, 
\label{energy2a}
\end{equation} 

\noindent which with the aid of equations \eqref{thermo} and \eqref{entropy},
gives following expression for the entropic force:

\begin{equation}
 F = \frac{GmM}{\xi x^\alpha r^2}.
\label{force2}
\end{equation} 

\noindent  The choice \( \alpha = 0 \) and \( \xi = 1 \) results into
Newton's law of gravity and \( \alpha = 1 \) with \( \xi = 1 \) converge to
MOND's force formula~\eqref{force1a}.

  The acceleration exerted on the test mass \( m \) is then given by:

\begin{equation}
 a = \frac{a_0}{r^2}\frac{GM}{a_0}\frac{1}{x^\alpha} = a_0
   \frac{x^2}{x^\alpha}.
\label{accel_thermo}
\end{equation} 

\noindent It is convenient to express it in this form, since
we can easily compare it with \eqref{accel_mendoza}.
It can be observed that given the form proposed for the energy $E$, in the
Newtonian regime, the ratio $x^2/x^\alpha \rightarrow x^2$, and in the modified
one, $x^2/x^\alpha \rightarrow x$, which are the same limits satisfied by
the function $f(x)$, as discussed in section \ref{introduction}.   Since
the transition function \( f(x) \) is given by equation~\eqref{f_series} it
follows that the equipartition energy can be written in a very general form as:

\begin{equation}
  E = \frac{1}{2}Nk_\textrm{B} T   x \frac{1\pm x^{n}}{1\pm x^{n+1}}
\label{energy2mas}
\end{equation}

\subsection*{Case (C)}
\label{subsection3}

Finally, we study the case for which \( \alpha = \beta = 0 \) in
equation~\eqref{energy_general-2} is considered. Equating the resulting
equation to \( M c^2 \) it is possible to find the temperature in terms
of the mass and distance and so with the aid of relation~\eqref{thermo}
it follows that:

\begin{equation}
 F = G\frac{mM}{\xi r^2} \left(\frac{a_0r}{c^2}\right)^\gamma.
\label{}
\end{equation} 

 The choice $\gamma=0$ and \( \xi = 1 \) yields Newton's gravitational law.  
A MOND-like limit
\( 1 / r \) seems to be possible when \( \gamma = 1 \), but the square root
dependence on the mass cannot be achieved with this choice.  Furthermore,
the speed of light still appears on this non-relativistic limit.
This all means that the dimensionless parameter \( \lambda  / r \) must
not appear on equation~\eqref{energy_general-2}, which means that \( \gamma
= 0 \).

  To summarise this Section, note that the inclusion of Milgrom's
acceleration constant as a fundamental quantity of nature related to
gravitational phenomena, is capable of generalising the equipartition law
in such a way that either Newton or MOND force formulae can be obtained.
This result reinforces the idea that, at all scales, gravity is an
emergent force with a thermodynamic nature.

\section{Modifications to the holographic principle}
\label{mod_information}
\label{sec:MLE}

 \noindent
 Let us now search for a MOND-like force formula assuming that the
equipartition law has its usual form and allowing for modifications related
to the way in which information is stored on the holographic screens (cf.
equation~\eqref{nbits}).  The only additional constant that needs to be
taken into consideration is \( a_0 \) and so, Buckingham's theorem of
dimensional analysis requires that the number of bits \( N \) stored on a
screen is given by:

\begin{equation}
 N = \frac{4\pi r^2}{l_\textrm{P}^2}g(x),
\label{nbits_mod}
\end{equation} 

\noindent where $g(x)$ is an unknown dimensionless function.  Since
equation~\eqref{nbits_mod} is a generalisation of the standard
relation~\eqref{nbits}, the function \( g(x) \rightarrow 1 \) in the
Newtonian regime of gravity, i.e. for \( x \gg 1 \).  
The previous equation assumes the validity
of the holographic principle and only changes the quantitative 
way in which information is stored on a particular screen.   Let us assume
that the function \( g(x) \) is a power law, i.e.  \( g(x) = x^\zeta \),
to obtain:

\begin{equation}
  N = \frac{ 4\pi r^2 }{ l_\text{P}^2 } x^\zeta,
\label{enemas}
\end{equation}

 With this in mind, we can follow Verlinde's analysis. Assuming the validity
of the equipartition law of energy, and the equivalence between mass and energy
inside the screen, along with equation \eqref{enemas}, it follows that:

\begin{equation}
 T =  \frac{GM \hbar}{2\pi ck_B r^2 x^\zeta}.
\label{temperature}
\end{equation} 

\noindent Direct substitution of the previous equation on
relation~\eqref{entropy}, with the aid of the first law of
thermodynamics~\eqref{thermo} yields the following expression for the
acceleration caused by this entropic force:

\begin{equation}
 a = \frac{GM}{r^2 x^\zeta}.
\label{accel2}
\end{equation} 

\noindent The choice \( \zeta = 1 \) yields a MONDian gravitational force
and \( \zeta = 0 \) a Newtonian one.  Comparison of the previous equation
with~\eqref{accel_mendoza} it follows that the complete
transition~\eqref{nbits_mod} is given by:

\begin{equation}
 N = \left( \frac{4\pi r^2}{l_\textrm{P}^2} \right) \, x 
   \frac{1\pm x^n}{1\pm x^{n+1}}.
\label{g_series}
\end{equation} 

  Finally, a possible alternative way to modify Verlinde's result it to
combine these two approaches, i.e. assume that both the equipartition law
of energy and the holographic principle get modified when MONDian effects are
introduced. Based on the previous analysis, this can be studied if we introduce
the parameter $x$ as follows:

\begin{equation}
 E = \frac{1}{2} N k_\text{B} T p(x), \quad 
   N = \frac{4\pi r^2}{l_\textrm{P}^2}q(x),
\label{p_q}
\end{equation}

\noindent where $p(x)$ and $q(x)$ are unknown functions of $x$.  Given
that both $p$ and $q$ are proportional to $E$ and $N$ respectively, they
will be inversely proportional  to the gravitational acceleration, i.e.:

\begin{equation}
 a = \frac{GM}{r^2}\frac{1}{p(x)q(x)} = a_0\frac{x^2}{p(x)q(x)},
\label{accel_p_q}
\end{equation}

\noindent and so,

\begin{equation}
  p(x) \, q(x) = \frac{ x^2 }{ f(x) },
\end{equation}

\noindent according to equation~\eqref{accel_mendoza} with \( p(x) \) and
\( q(x) \) tending to \( 1 \) in the Newtonian limit, i.e. when \( x \gg 1
\).  Also, the previous equation imposes the following restriction: \( p(x)
g(x) \rightarrow x \) in the MONDian regime of gravity, i.e. \( x \ll 1 \)

\section{Discussion}
\label{discussion}

 \noindent
 As explained by \cite{famaey}, \cite{mendoza12} and in a more profound
and empirical way by \cite{mendoza_olmo}, if gravitational phenomena
requires to be modified at a certain scales of mass and length one
needs to incorporate a new fundamental constant of nature relevant to all
gravitational phenomena at those scales.  
This gravitational constant is as important as Newton's constant of
gravity and can be mathematically manipulated as to have dimensions of
acceleration which converge to Milgrom's acceleration constant \( a_0 \).
This is so since gravitational phenomena does not follow the standard
Newtonian (or general relativistic) behaviour of gravity at scales which
greatly differ from the ones in which precise gravitational experiments
have been performed to test the validity of Newton's law of gravitation
(or Einstein's general relativity -cf.~\cite{will93}). The behaviour
of gravity at those scales can be considered as independent of the
behaviour of standard gravity and as such a new fundamental constant
of nature has to be introduced into the description of gravitational
phenomena, which is a standard procedure to follow when extensions of a
particular physical theory are performed (e.g. \cite{sedov}).

  In this article we have introduced this extra fundamental constant of
nature \( a_0 \) in the description of gravity and used thermodynamic
and information properties of space and time in order to show that a
MONDian force law can be obtained by assuming the validity of the 
holographic principle. Specifically, this has been studied under two
different approaches using Buckingham's \( \Pi \) theorem of dimensional
analysis:

  (1) In the first approach, the equipartition law is modified and the
holographic principle keeps its standard form, resulting on a
temperature scale \( T_* \) (corresponding to the Unruh temperature evaluated at
the Milgrom acceleration \( a_0 \)).
This temperature corresponds to the one already found by 
\cite{debye_article} who studied the problem of the emergence of
gravity modifying the equipartition law using a Debye model. \cite{pazy}
also worked on modifications focused on thermodynamical statistical
properties  and in the same way as \cite{debye_article} dynamical
modifications with its corresponding MOND transition function \( \mu(a/a_0)
\) were obtained.  As explained by \cite{mendoza11}, direct dimensional
analysis points towards a modification of the gravitational force, and the
calculations we performed in this work were done by keeping Newton's second law
untouched,  allowing for an extension on the gravitational sector.

  (2) In a second approach, modifications to the holographic principle
 -with the equipartition law unchanged- have showed to be able to explain
in a natural way how gravity transits from a Newtonian regime to a
modified one.

  The important point about these two different ways is that they are
consistent with the general formula for acceleration experienced by a
particle under a gravitational field given in \eqref{accel_mendoza}.
In this sense, and in the context of emergent gravity, it suggests that
the transition observed across different astrophysical systems could be
a consequence of a modification at a deeper level in the equipartition
law and/or in the holographic principle, i.e the observed effects at
large scales in gravitational systems reflect the behaviour
of physical laws at a deeper thermodynamical level. More generally,
it has been also considered the possibility of modifications of both
the holographic principle and the equipartition law in such regimes.

  As pointed by \cite{hernandez12}, observations of
globular clusters yield a lower limit on the exponent \( n \gtrsim 8 \) in
equation~\eqref{f_series}, meaning that the transition function \( f(x) \)
is quite abrupt, i.e. \( f(x) = x \) for \( x \leq 1 \) (MONDian regime) 
and \( f(x) = x^2 \) for \( x \geq 1 \) (Newtonian regime), with almost no
soft transition from one regime to the other.  This means that the
transition functions \( E \), \( N \), \( p(x) \) and \( q(x) \)
calculated in this article must present a rather abrupt transition.

  A full non-relativistic theory of gravity can be constructed assuming a
modification of inertia as described by \cite{famaey}, but as shown in
this work the modification naturally appears in the force sector and not
on the dynamical one.  

 With a few natural assumptions about space and information, the main
result of this article is to show that gravity can be considered an
emergent phenomenon also in the MONDian regime.  This suggests that the
force of gravity on this extended regime is not a fundamental force of
nature, but a consequence of the inherent properties of space and time.
Since \cite{verlinde} showed that Newtonian gravity emerges from the
thermodynamic properties of space and time, this all suggests that
gravitation is an emergent phenomenon at all scales of mass and length.

\section{Acknowledgements}
\label{acknowledgements}

  We thank an anonymous referee for his fruitful comments on the first
version of this article.  This work was supported by a DGAPA-UNAM grant
(PAPIIT IN111513-3) and a CONACyT grant (240512). DAC and SM thank
support granted by CONACyT 480147 and 26344.  The authors gratefully
acknowledge Ehoud Pazy and Hristu Culetu for the
valuable comments made of an earlier version of this article.

\bibliographystyle{aipauth4-1}
\bibliography{thermomondian}

\end{document}